\documentclass[a4paper,notitlepage,twocolumn,footnoteinbib,longbibliography,superscriptaddress,10pt]{revtex4-1}

\usepackage{amsmath,amssymb,amsthm,mathtools,color,listings,graphicx}
\usepackage[colorlinks=true,linkcolor=blue,citecolor=blue]{hyperref}
\DeclareMathAlphabet{\mathbbb}{U}{bbold}{m}{n}

\begin{document}

\title{Dawn and fall of non-Gaussianity in the quantum parametric oscillator}
\author{Marcello Calvanese Strinati}
\email{marcello.calvanesestrinati@gmail.com}
\affiliation{Centro Ricerche Enrico Fermi (CREF), Via Panisperna 89a, 00184 Rome, Italy}
\author{Claudio Conti}
\affiliation{Physics Department, Sapienza University of Rome, 00185 Rome, Italy}
\affiliation{Centro Ricerche Enrico Fermi (CREF), Via Panisperna 89a, 00184 Rome, Italy}
\date{\today}

\begin{abstract}
Systems of coupled optical parametric oscillators (OPOs) forming an Ising machine are emerging as large-scale simulators of the Ising model. The advances in computer science and nonlinear optics have triggered not only the physical realization of hybrid (electro-optical) or all-optical Ising machines, but also the demonstration of quantum-inspired algorithms boosting their performances. To date, the use of the quantum nature of parametrically generated light as a further resource for computation represents a major open issue. A key quantum feature is the non-Gaussian character of the system state across the oscillation threshold. In this paper, we perform an extensive analysis of the emergence of non-Gaussianity in the single quantum OPO with an applied external field. We model the OPO by a Lindblad master equation, which is numerically solved by an \emph{ab initio} method based on exact diagonalization. Non-Gaussianity is quantified by means of three different metrics: Hilbert-Schmidt distance, quantum relative entropy, and photon distribution. Our findings reveal a nontrivial interplay between parametric drive and applied field: (i) Increasing pump monotonously enhances non-Gaussianity, and (ii) Increasing field first sharpens non-Gaussianity, and then restores the Gaussian character of the state when above a threshold value.
\end{abstract}

\maketitle

\section{Introduction}
Hard optimization problems are permeating several areas of modern science and society. Their rapidly-increasing computational complexity nowadays pairs with the evident limits of conventional computer architectures, fostering the investigation of innovative specialized paradigms and devices. In this respect, optical systems are emerging as promising alternative computing platforms~\cite{petermchannon2023}. Leveraging the mapping of complex optimization to Ising Hamiltonians~\cite{10.3389/fphy.2014.00005}, the quest of solving a large class of problems translates into building a system capable of simulating the classical Ising model and efficiently finding its lowest-energy configuration.

Specifically, systems of optical parametric oscillators (OPOs) have emerged as a valuable platform to solve the Ising model. When pumped by an external drive above the oscillation threshold, an OPO undergoes phase-dependent amplification forcing the phase of the optically amplified signal to be either $0$ or $\pi$ with respect to the phase of the pump. These two states simulate the ``spin-up'' and ``spin-down'' configurations of a classical Ising spin. This circumstance is behind the use of networks of coupled OPOs as computing machines (called Ising machines) to find the ground state of the classical Ising model~\cite{yamamoto2020isingmachine}. Recently, a significant effort has been put in exploiting classical properties of OPOs to enhance computational speed and efficiency~\cite{doi:10.1002/qute.202000045,leleuchaoticamplitudecontro2021,hgoto2021cim,strinati2022hyperspinmachine,arXiv:2308.02329}.

The question on whether quantum features of the parametrically generated light can be employed to further boost OPO-based computing machines has also been raised~\cite{PhysRevA.102.062419,yamamoto2021nopo,PhysRevA.106.022409}. However, no clear answer is available to date because of the difficulty in the analytical and numerical description of quantum OPO networks compared to their classical counterparts. A major issue is the identification of specific quantum properties that can enhance the computational performance.

One of such quantum features is the non-Gaussian nature of the state~\cite{PRXQuantum.2.030204}, which is the focus of this work. Previous work discussed the presence of non-Gaussianity in OPOs close to the threshold~\cite{DAuria:05,PhysRevA.81.033846} by observing non-Gaussian statistics in the photon distribution. The emergence of non-Gaussian correlations as the system is driven above the oscillation threshold is one of the key features that is envisioned to improve quantum tunneling during the quantum parallel search~\cite{s41534-017-0048-9} and thus enhance the OPO-based Ising machines. However, a systematic study on the way non-Gaussianity emerges is missing.

In this work, we report on an extensive analysis of non-Gaussianity in the quantum OPO in different parameter regimes. We model the OPO by a driven-dissipative open quantum system described by a Lindblad master equation accounting for two-photon gain (pump) and subject to one- and two-photon dissipation (intrinsic loss and pump saturation, respectively). We numerically obtain the full density matrix of the system by resorting to an \emph{ab-initio} method, by projection of the master equation on the Fock (number) basis and subsequent exact diagonalization of the Liouvillian superoperator~\cite{PhysRevA.98.042118}. Non-Gaussianity is first quantified as a function of the pump amplitude by comparing three different metrics: Hilbert-Schmidt distance~\cite{doi:10.1080/09500340008233385,PhysRevA.76.042327,PhysRevA.82.052341}, quantum relative entropy $s(\hat\rho)$~\cite{PhysRevA.82.052341,PhysRevA.78.060303,PhysRevA.88.012322}, and photon distribution~\cite{DAuria:05}. Then, non-Gaussianity is studied in the presence of a one-photon drive (additive field) by computing the quantum relative entropy as a function of both pump amplitude and applied field strength.

We find that, while the quantum state is well described by a Gaussian state for sufficiently low pump, non-Gaussianity dominates above threshold. Specifically, increasing pump causes a monotonous growth of non-Gaussianity, while increasing additive field first makes non-Gaussianity to grow, and then causes a steep decrease, suggesting a restoration of the Gaussian nature of the state for large additive field.

This paper is organized as follows: In Sec.~\ref{sec:themodel1} we introduce the quantum model of the OPO and review the corresponding classical model. In Sec.~\ref{sec:masterequationinthefockbasis1}, we discuss our numerical procedure, first addressing the case of zero additive field. We present our numerical results on the Wigner function in Sec.~\ref{sec:wignerfunction1}. The measurements of non-Gaussianity are discussed in Sec.~\ref{sec:measirementsnongaussianity1}, and their analysis is extended to the case of nonzero additive field in Sec.~\ref{sec:inclustionofadditivefield1}. We draw our conclusions in Sec.~\ref{sec:perspectives1}, and report additional analytical and numerical details in the appendices.

\section{The model}
\label{sec:themodel1}
In this section, we introduce our model of the quantum optical parametric oscillator and review for the sake of completeness its main properties in the classical (mean-field) limit.
\subsection{Quantum master equation}
\label{sec:masterequation1}
We model the OPO as a driven-dissipative open quantum system described by a density operator $\hat\rho$, obeying the following master equation ($\hbar=1$)~\cite{,breuer2002theory,carmichael2010optics}
\begin{equation}
\frac{d}{dt}\hat\rho=\mathcal{L}\hat\rho(t)=\frac{1}{i}\left[\hat H_0,\hat\rho\right]+\mathcal{D}_{{\rm 1ph}}\left(\hat\rho\right)+\mathcal{D}_{{\rm 2ph}}\left(\hat\rho\right) \,\, ,
\label{eq:masterequationquantumopos1}
\end{equation}
where $\mathcal{L}$ is the Liouvillian superoperator. In Eq.~\eqref{eq:masterequationquantumopos1}, we define
\begin{equation}
\hat H_0=i\frac{h}{8}\left({(\hat a^\dag)}^2-\hat a^2\right) \,\, ,
\label{eq:masterequationquantumopos2}
\end{equation}
as the Hamiltonian describing two-photon gain (parametric amplification) by a real field of amplitude $h>0$, and
\begin{subequations}
\begin{align}
\mathcal{D}_{\rm 1ph}\left(\hat\rho\right)&=g\left(\hat a\,\hat\rho\,\hat a^\dag-\frac{1}{2}\left\{\hat a^\dag\hat a,\hat\rho\right\}\right)\\
\mathcal{D}_{\rm 2ph}\left(\hat\rho\right)&=\frac{\beta}{2}\left(\hat a^2\,\hat\rho\,{(\hat a^\dag)}^2-\frac{1}{2}\left\{{(\hat a^\dag)}^2\hat a^2,\hat\rho\right\}\right) \,\, ,
\end{align}
\label{eq:masterequationquantumopos3}
\end{subequations}
are the dissipators representing one- and two-photon losses. These processes describe the intrinsic cavity loss (quantified by $g>0$) and the nonlinear saturation (quantified by $\beta>0$), respectively. In Eqs.~\eqref{eq:masterequationquantumopos2} and~\eqref{eq:masterequationquantumopos3}, $\hat a$ ($\hat a^\dag$) is the photon annihilation (creation) operator, obeying the bosonic commutation relations $[\hat a,\hat a^\dag]=1$ and $[\hat a,\hat a]=0$.

\subsection{Classical limit}
From Eq.~\eqref{eq:masterequationquantumopos1}, we obtain the equation of motion for $\hat a$ by the adjoint master equation
\begin{equation}
\frac{d\hat a}{dt}=\frac{h}{4}\hat a^\dag-\frac{g}{2}\hat a-\frac{\beta}{2}\hat a^\dag\hat a^2 \,\, .
\label{eq:masterequationquantumopos4}
\end{equation}
By taking the mean-field approximation $\hat a\rightarrow\langle\hat a\rangle\equiv A$ in Eq.~\eqref{eq:masterequationquantumopos4}, the classical equations of motion describing the dynamics of the complex OPO amplitude $A$ are obtained~\cite{PhysRevA.100.023835,goto2019kpoandopo}
\begin{equation}
\frac{dA}{dt}=\frac{h}{4}A^*-\frac{1}{2}\left(g+\beta{|A|}^2\right)A \,\, .
\label{eq:masterequationquantumopos5}
\end{equation}
When the pump amplitude $h$ is below the classical oscillation threshold value $h_{\rm th}=2g$, the dynamics in Eq.~\eqref{eq:masterequationquantumopos5} suppresses both the real and imaginary part of $A$ (respectively ${\rm Re}[A]$ and ${\rm Im}[A]$). The only fixed point of the dynamics (defined by the condition $d\bar{A}/dt=0$, where the overline denotes the steady-state value) is the origin of the complex plane, i.e., ${\rm Re}[\bar{A}]={\rm Im}[\bar A]=0$. When the pump amplitude is driven above threshold ($h>h_{\rm th}$), the origin becomes a saddle point, giving raise to two symmetric stable fixed points on the real axis by a pitchfork bifurcation~\cite{strogatz2007nonlinear}. The amplitude at these nontrivial fixed points from Eq.~\eqref{eq:masterequationquantumopos5} is readily found
\begin{equation}
{\rm Re}\left[\bar{A}\right]=\pm\sqrt{\frac{1}{\beta}\left(\frac{h}{2}-g\right)} \qquad {\rm Im}\left[\bar{A}\right]=0 \,\, .
\label{eq:masterequationquantumopos6}
\end{equation}
Above threshold, the system converges to the fixed point in Eq.~\eqref{eq:masterequationquantumopos6} with sign determined by the initial condition $A(t=0)$, a phenomenology that is reminiscent of the spontaneous $\mathbb{Z}_2$ (Ising) symmetry breaking.

\section{Master equation in the Fock basis}
\label{sec:masterequationinthefockbasis1}
We now discuss the numerical solution of the quantum master equation in Eq.~\eqref{eq:masterequationquantumopos1}. Our goal is to find the exact density operator $\hat\rho$, from which any observable can be measured. To this end, we proceed by using an \emph{ab initio} method as follows. We choose the basis of Fock (number) states for the bosonic Hilbert space $\mathcal{H}={\rm span}\{|n\rangle\}_{n=0}^{\infty}$ to represent $\hat\rho$ as a (infinite) real positive definite matrix with elements $\rho_{mn}\equiv\langle m|\hat\rho|n\rangle$, so that
\begin{equation}
\hat\rho=\sum_{m,n=0}^{\infty}\rho_{mn}\,|m\rangle\langle n| \,\, .
\label{eq:densitymatrixfockrepresentation1}
\end{equation}
The projection of Eq.~\eqref{eq:masterequationquantumopos1} onto the Fock states allows to obtain the equations of motion for all the elements $\rho_{mn}$ in the following tensor form
\begin{equation}
\frac{d}{dt}\rho_{mn}=\sum_{r,s=0}^{\infty}\mathcal{L}^{rs}_{mn}\,\rho_{rs} \,\, ,
\label{eq:masterequationquantumopos7}
\end{equation}
where the nonzero elements of the Liouvillian tensor $\mathcal{L}^{rs}_{mn}$ are the projected right-hand side of Eq.~\eqref{eq:masterequationquantumopos1} and are reported in Appendix~\ref{appendix:liuvilliantensorinthefockbasis1}.

While in general the Fock states are upper unbounded, in our numerics, we truncate the Hilbert space up to $n_{\rm max}-1$ particles, i.e., $\mathcal{H}={\rm span}\{|n\rangle\}_{n=0}^{n_{\rm max}-1}$, in order to represent operators (superoperators) as matrices (tensors) of finite size~\cite{PhysRevA.43.6194}. In particular, $\rho_{mn}$ and $\mathcal{L}^{rs}_{mn}$ are a $n_{\rm max}\times n_{\rm max}$ matrix and a $n_{\rm max}\times n_{\rm max}\times n_{\rm max}\times n_{\rm max}$ tensor, respectively. The steady state density matrix $\bar{\rho}_{mn}$, found from Eq.~\eqref{eq:masterequationquantumopos1} as customary by imposing $d\bar{\rho}/dt=\mathcal{L}\bar{\rho}=0$, is obtained by the exact diagonalization of $\mathcal{L}^{rs}_{mn}$, reshaped as a matrix, as the eigenvector of the Liouvillian associated to the zero eigenvalue~\cite{PhysRevA.94.033841,PhysRevA.98.042118}.

Physically, the truncation of the Hilbert space is possible thanks to the presence of the nonlinear saturation dissipator $\mathcal{D}_{{\rm 2ph}}(\hat\rho)$ in Eq.~\eqref{eq:masterequationquantumopos3}, which naturally sets an upper bound for the average number of photons $\langle\hat a^\dag\hat a\rangle$ in the system that is approximatively given by the squared classical fixed-point amplitude in see Eq.~\eqref{eq:masterequationquantumopos6}: $\langle\hat a^\dag\hat a\rangle\simeq(h/2-g)/\beta$. Therefore, to have a faithful representation of $\hat\rho$ on the truncated Hilbert space, it is sufficient to choose $n_{\rm max}$ such that $|\rho_{mn}|<\epsilon$ with $\epsilon$ vanishingly small, for all $m,n>n_{\rm max}$. We checked that this condition is ensured for $\hat\rho$ in all our numerical simulations.

\begin{figure}[t]
\centering
\includegraphics[width=8.4cm]{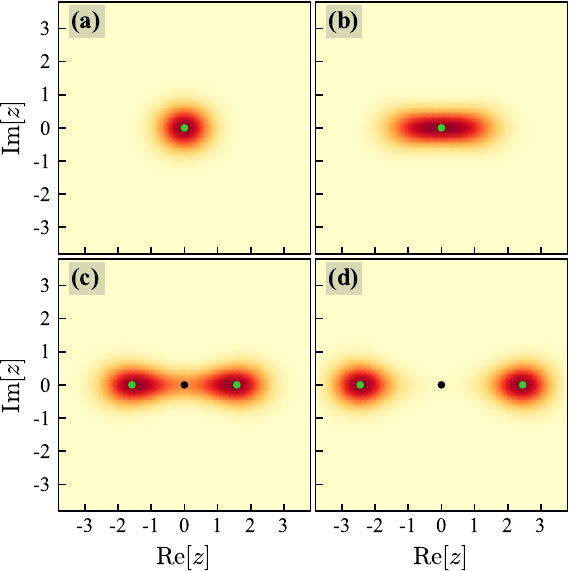}
\caption{Colormap of the Wigner function $W(z)$, computed as in Eq.~\eqref{eq:wignerfunctiondefinition3}, in the ${\rm Im}[z]$ vs. ${\rm Re}[z]$ plane, for different values of the pump amplitude $h$: \textbf{(a)} $h=0.2$ ($80\%$ below threshold), \textbf{(b)} $h=1.0$ (at threshold), \textbf{(c)} $h=1.5$ ($50\%$ above threshold), and \textbf{(d)} $h=2.2$ ($120\%$ above threshold). Color coding: Yellow for $W(z)=0$, and other colors for $W(z)>0$. Numerical parameters are $n_{\rm max}=40$, $g=0.5$, and $\beta=0.1$. The green and black points are the stable and unstable fixed points of the classical equations of motion in Eq.~\eqref{eq:masterequationquantumopos6}, respectively. The Wigner function is very close to one Gaussian lobe far below threshold. As $h$ approaches the threshold, it becomes an elongated cigar-shaped lobe, which eventually splits into two lobes on the real axis, symmetric with respect to ${\rm Re}[z]=0$.}
\label{fig:wignerfunctioncomputation1}
\end{figure}

\section{Wigner function}
\label{sec:wignerfunction1}
A useful observable that can be measured from the numerically obtained density matrix $\rho_{mn}$ is the Wigner quasi-probability distribution function $W(z)$, which provides a representation of the quantum state in the complex quadrature space $z=(X+iP)/\sqrt{2}$, where $X$ and $P$ are the position and momentum coordinates, respectively. The Wigner function is defined as the complex Fourier transform of the characteristic function $\chi(\xi)={\rm Tr}[\hat D_\xi\hat\rho]$, where $\hat D_\xi=e^{\xi\hat a^\dag-\xi^*\hat a}$ is the displacement operator, i.e., $W(z)=\frac{1}{\pi^2}\int_{\mathbb{C}}d^2\xi\,e^{z\xi^*-z^*\xi}\,\chi(\xi)$, where the integral extends over the complex plane. Using a series of identities, one can show that the Wigner function is equivalently rewritten as~\cite{PhysRev.177.1857,PhysRev.177.1882}
\begin{equation}
W(z)=\frac{2}{\pi}\,{\rm Tr}\left[\hat D_{2z}\,e^{i\pi\,\hat a^\dag\hat a}\hat\rho\right] \,\, .
\label{eq:wignerfunctiondefinition2}
\end{equation}
Equation~\eqref{eq:wignerfunctiondefinition2} is particularly useful when $\hat\rho$ is represented in the Fock basis. Indeed, by using the resolution of the identity $\sum_{n=0}^{\infty}|n\rangle\langle n|=\hat{\mathbbb{1}}$, one has
\begin{equation}
W(z)=\frac{2}{\pi}\sum_{m,n}{(-1)}^m\langle n|\hat D_{2z}|m\rangle\,\rho_{mn} \,\, .
\label{eq:wignerfunctiondefinition3}
\end{equation}
The matrix representation of the displacement operator in the Fock basis $\langle n|\hat D_{z}|m\rangle$ is reported in Appendix~\ref{appendix:displaementoperatorfockbasis1}.

The numerical results on the Wigner function are shown in Fig.~\ref{fig:wignerfunctioncomputation1}, where we plot $W(z)$ as a colomap in the ${\rm Im}[z]$ vs. ${\rm Re}[z]$ plane for different values of the pump amplitude $h$, relative to the classical oscillation threshold $h_{\rm th}=2g$, on which we overlap the classical fixed points from Eq.~\eqref{eq:masterequationquantumopos6} as green and black dots for stable and unstable points, respectively. We see that below threshold, the Wigner function consists of one lobe centered about the stable origin, and it develops a symmetric two-lobe structure on the real axis around the origin (which becomes a saddle after the pitchfork bifurcation) as the pump amplitude is driven above the classical threshold. From the quantum point of view, such a symmetric Wigner function signifies that the state is found on each lobe with equal probability. The corresponding classical behaviour is explained by the fact that the two stable fixed points are equally attractive, i.e., their basins of attraction are of equal size so that the probability to converge to either fixed point is the same for random initial conditions close to the origin.

It is known that the quantum state or a sub-threshold OPO is a squeezed state~\cite{walls1981,Wu:87}, which is a Gaussian state (in particular for zero pump the system is in the vacuum state). Instead, the two-lobe structure of $W(z)$ is a clear indication of the non-Gaussian nature of the state above threshold. Specifically, far above threshold $W(z)$ resembles two symmetric Gaussian lobes, suggesting that the quantum state tends to a mixture of coherent states: $\hat\rho\simeq(|\alpha\rangle\langle\alpha|+|-\alpha\rangle\langle-\alpha|)/2$ with $|\alpha|\simeq\bar{A}$ in Eq.~\eqref{eq:masterequationquantumopos6}, which is indeed non-Gaussian~\cite{PRXQuantum.2.030204}. A natural question therefore arises: How does non-Gaussianity emerges from the Gaussian state as the pump amplitude is driven from below to above threshold?

\section{Measurements of non-Gaussianity}
\label{sec:measirementsnongaussianity1}
In this section, we quantify the deviation from Gaussianity of the quantum state $\hat\rho$ as the system crosses the oscillation threshold by comparing three different metrics: The degree of non-Gaussianity $\delta(\hat\rho)$ based on the Hilbert-Schmidt distance~\cite{doi:10.1080/09500340008233385,PhysRevA.76.042327,PhysRevA.82.052341}, the quantum relative entropy $s(\hat\rho)$~\cite{PhysRevA.82.052341,PhysRevA.78.060303,PhysRevA.88.012322}, and non-Gaussianity $Q(\hat\rho)$ extracted from the photon distribution of $\hat\rho$. All these metrics quantify the deviation of the actual quantum state $\hat\rho$ from a reference state $\hat\tau$ defined as the Gaussian state having the same first and second moments (covariance matrix) of $\hat\rho$. Since $\hat\tau$ is Gaussian, the determination of the first moments and covariance matrix of $\hat\rho$ fully determines $\hat\tau$.

\subsection{Determination of the Gaussian reference state}
\label{sec:determinationofthegaussianreferencestate1}
We first discuss how the state $\hat\tau$ is defined. Let us denote by $\hat{\mathbf{R}}=(\hat X,\hat P)$ the vector of the two quadratures $\hat R_1\equiv\hat X=(\hat a+\hat a^\dag)/\sqrt{2}$ and $\hat R_2\equiv\hat P=(\hat a-\hat a^\dag)/i\sqrt{2}$. From the state $\hat\rho$, the first moments $\langle\hat{\mathbf{R}}\rangle$ and covariance matrix $\mathbf{\Sigma}$ (which is a $2\times2$ real and symmetric matrix) are found as customary as
\begin{equation}
\left\langle\hat{R}_j\right\rangle={\rm Tr}\left[\hat{R}_j\,\hat\rho\right] \qquad \Sigma_{jk}=\frac{1}{2}{\rm Tr}\left[\left\{\Delta\hat R_j,\Delta\hat R_k\right\}\hat\rho\right] \,\, ,
\label{eq:firstandsecondmomentsrho1}
\end{equation}
where $\Delta\hat R_j=\hat R_j-\langle\hat R_j\rangle$. From our numerical simulations, the first moments and covariance matrix of $\hat\rho$ are readily computed by plugging in Eq.~\eqref{eq:firstandsecondmomentsrho1} the Fock state expansion in Eq.~\eqref{eq:densitymatrixfockrepresentation1} with $\rho_{mn}$ computed as explained before, and by recalling that $\hat a|n\rangle=\sqrt{n}|n-1\rangle$ and $\hat a^\dag|n\rangle=\sqrt{n+1}|n+1\rangle$. Let us observe that, due to $\mathbb{Z}_2$ symmetry, which translates in phase space as invariance under inversion symmetry $\hat{\mathbf{R}}\rightarrow-\hat{\mathbf{R}}$, the first moments in our case are zero, and thus the computation of the covariance matrix simplifies to $\Sigma_{jk}=\frac{1}{2}{\rm Tr}[\{\hat R_j,\hat R_k\}\hat\rho]$.

As said before, the computed first moments and covariance matrix of $\hat\rho$ are by construction the same of $\hat\tau$. Since the generic single-mode Gaussian state is given by the displaced squeezed thermal state~\cite{PhysRevA.76.042327}
\begin{equation}
\hat\tau=\hat D_\alpha\,\hat S(\xi)\,\hat\rho_{\rm th}(\overline{n})\,\hat S^\dag(\xi)\,\hat D^\dag_\alpha \,\, ,
\label{eq:firstandsecondmomentsrhogaussianstate1}
\end{equation}
with complex $\alpha$ and $\xi$, and $\overline{n}\geq0$, where the squeezing operator is $\hat S(\xi)=e^{\left(\xi^*\,\hat a\hat a-\xi\,\hat a^\dag\hat a^\dag\right)/2}$ and the thermal state with average number of thermal particles $\overline{n}$ is
\begin{equation}
\hat\rho_{\rm th}(\overline{n})=\sum_{n=0}^{\infty}f_n|n\rangle\langle n| \qquad f_n=\frac{\overline{n}^n}{{(\overline{n}+1)}^{n+1}} \,\, ,
\label{eq:fockbasisexpansionthermalstate1}
\end{equation}
the state $\hat\tau$ is determined by finding $\alpha$, $\xi$, and $\overline{n}$ from $\langle\hat X\rangle$, $\langle\hat P\rangle$, and $\mathbf{\Sigma}$ of $\hat\rho$.

The displacement $\alpha$ affects the first moments only, and one has ${\rm Re}[\alpha]=\langle\hat X\rangle/\sqrt{2}$ and ${\rm Im}[\alpha]=\langle\hat P\rangle/\sqrt{2}$. In our case, since the first moments are zero, one readily has $\alpha=0$ and thus $\hat D_\alpha=\hat{\mathbbb{1}}$. Instead, the squeezing $\xi$ and thermal number of photons $\overline{n}$ affect covariance matrix only, whose form is reviewed in Appendix~\ref{appendix:covariancematrixsqueezedthermalstate1}. From our numerical simulations, we observe that the covariance matrix $\mathbf{\Sigma}$ of $\hat\rho$ (and thus of $\hat\tau$) is a diagonal matrix with $\Sigma_{11}>\Sigma_{22}$. From Appendix~\ref{appendix:covariancematrixsqueezedthermalstate1} it follows that $\hat\tau$ is defined with real $\xi$ and $\overline{n}$ given by
\begin{equation}
\xi=-\frac{1}{4}\log\left(\frac{\Sigma_{11}}{\Sigma_{22}}\right) \qquad \overline{n}=\sqrt{\Sigma_{11}\Sigma_{22}}-\frac{1}{2} \,\, .
\label{eq:xiandngaussianstatetatu1}
\end{equation}
We recall that $\overline{n}$ is related to the symplectic eigenvalue $\nu$ of $\mathbf{\Sigma}$ by $\nu=\overline{n}+1/2=\sqrt{\Sigma_{11}\Sigma_{22}}=\sqrt{{\rm det}[\mathbf{\Sigma}]}$~\cite{serafini2017quantum}. The Fock representation of $\hat\tau$ in Eq.~\eqref{eq:firstandsecondmomentsrhogaussianstate1} with $\alpha=0$ is
\begin{equation}
\tau_{mn}=\sum_{v=0}^{\infty}f_v\,\langle m|\hat S(\xi)|v\rangle\langle v|\hat S^\dag(\xi)|n\rangle \,\, ,
\label{eq:gaussiantargetstate1}
\end{equation}
which, with $\xi$ and $\overline{n}$ in Eq.~\eqref{eq:xiandngaussianstatetatu1}, is a real and symmetric matrix, where $f_v$ is as in Eq.~\eqref{eq:fockbasisexpansionthermalstate1} and the expression of $\langle v|\hat S^\dag(\xi)|n\rangle=(\langle n|\hat S(\xi)|v\rangle)^*$ is reported in Appendix~\ref{appendix:squeezingoperatorfockbasis1}.

\subsection{Non-Gaussianity by Hilbert-Schimdt distance}
A natural way to quantify the deviation of $\hat\rho$ from Gaussianity is via the operator distance between $\hat\rho$ and $\hat\tau$ in the Hilbert-Schmidt metric~\cite{doi:10.1080/09500340008233385}
\begin{eqnarray}
D_{\rm HS}\left(\hat\rho,\hat\tau\right)&=&\sqrt{{\rm Tr}\left[{\left(\hat\rho-\hat\tau\right)}^2\right]}\nonumber\\
&=&\sqrt{{\rm Tr}\left[\hat\rho^2\right]+{\rm Tr}\left[\hat\tau^2\right]-2\,{\rm Tr}\left[\hat\rho\,\hat\tau\right]} \,\, ,
\label{eq:tracerhorhogaussian1}
\end{eqnarray}
where the purity of $\hat\rho$ in Eq.~\eqref{eq:densitymatrixfockrepresentation1} and of $\hat\tau$ in Eq.~\eqref{eq:firstandsecondmomentsrhogaussianstate1} are (see also Appendix~\ref{appendix:covariancematrixsqueezedthermalstate1})
\begin{equation}
{\rm Tr}\left[\hat\rho^2\right]=\sum_{m,n=0}^{\infty}\rho_{mn}^2 \qquad {\rm Tr}\left[\hat\tau^2\right]=\frac{1}{2\overline{n}+1} \,\, .
\label{eq:tracerhorhogaussian2}
\end{equation}
Moreover
\begin{equation}
{\rm Tr}\left[\hat\rho\,\hat\tau\right]=\sum_{m,n=0}^{\infty}\rho_{mn}\,\tau_{mn} \,\, ,
\label{eq:scalarproductrhotauoverlap1}
\end{equation}
denotes the scalar product (overlap) between $\hat\rho$ and $\hat\tau$ (recall that both $\rho_{mn}$ and $\tau_{mn}$ are real and symmetric matrices). From Eq.~\eqref{eq:tracerhorhogaussian1}, the degree of non-Gaussianity is defined as~\cite{PhysRevA.76.042327,PhysRevA.82.052341}
\begin{equation}
\delta\left(\hat\rho\right)\coloneqq\frac{D^2_{\rm HS}(\hat\rho,\hat\tau)}{2\,{\rm Tr}[\hat\rho^2]} \,\, .
\label{eq:tracerhorhogaussian3}
\end{equation}
Notice that, in order to numerically compute the purities of $\hat\rho$ and $\hat\tau$, it is sufficient to determine $\hat\rho$ in the Fock basis, from which $\mathbf{\Sigma}$ and thus $\overline{n}$ in Eq.~\eqref{eq:xiandngaussianstatetatu1} are computed. Instead, the numerical computation of the overlap ${\rm Tr}\left[\hat\rho\,\hat\tau\right]$ requires also the Fock representation of $\hat\tau$ in Eq.~\eqref{eq:gaussiantargetstate1}.

\subsection{Quantum relative entropy}
Another observable that quantifies the non-Gaussian nature of the quantum state is provided by the quantum relative entropy between the actual state $\hat\rho$ and its Gaussian reference state $\hat\tau$~\cite{PhysRevA.82.052341,PhysRevA.78.060303}
\begin{equation}
s\left(\hat\rho\right)\coloneqq S\left(\hat\tau\right)-S\left(\hat\rho\right) \,\, ,
\label{eq:measurenongaussianityentropy1}
\end{equation}
where $S(\hat\rho)=-{\rm Tr}[\hat\rho\,\log(\hat\rho)]$ is the von Neumann entropy. For the state $\hat\rho$ in Eq.~\eqref{eq:densitymatrixfockrepresentation1}, the von Neumann entropy is defined in terms of the eigenvalues $\lambda_k\geq0$ of $\rho_{mn}$ as
\begin{equation}
S\left(\hat\rho\right)=-\sum_{k=0}^{\infty}\lambda_k\,\log\left(\lambda_k\right) \,\, .
\label{eq:measurenongaussianityentropy2}
\end{equation}
Instead, the von Neumann entropy of $\hat\tau$ readily follows from the diagonal representation of the thermal state in Eq.~\eqref{eq:fockbasisexpansionthermalstate1}, i.e., $S(\hat\tau)=-\sum_{n=0}^{\infty}f_n\log(f_n)$, which explicitly reads~\cite{PhysRevA.59.1820}
\begin{equation}
S\left(\hat\tau\right)=\left(\overline{n}+1\right)\log\left(\overline{n}+1\right)-\overline{n}\log\left(\overline{n}\right) \,\, .
\label{eq:measurenongaussianityentropy3}
\end{equation}
The fact that Eq.~\eqref{eq:measurenongaussianityentropy1} defines an exact distance-type measure of non-Gaussianity was shown in Ref.~\cite{PhysRevA.88.012322}.

\subsection{Euclidian distance between photon distributions}
While the degree of non-Gaussianity and quantum relative entropy in Eqs.~\eqref{eq:tracerhorhogaussian3} and~\eqref{eq:measurenongaussianityentropy1} provide exact measurements to quantify the non-Gaussian nature of the state, they require the full knowledge of the density matrix $\hat\rho$. However, reconstructing $\hat\rho$ requires complex state tomography techniques that are often unfeasible for large-dimensional systems~\cite{Toninelli:19}, hampering the experimental measurement of $\delta(\hat\rho)$ and $s(\hat\rho)$. To overcome this problem, we show that it is possible to obtain similar results as for Eq.~\eqref{eq:tracerhorhogaussian3} from solely the knowledge of the first moments, $\mathbf{\Sigma}$, and the photon distribution $p_n\simeq\rho_{nn}$. This fact has notable advantages in experiments, since first and second moments are measured by homodyne detection~\cite{orszag2016quantum}, while $\rho_{nn}$ is measured by photon counting~\cite{paris2004quantum}.

The measured $\mathbf{\Sigma}$ of the full quantum state $\hat\rho$ is used to determine the Gaussian target $\hat\tau$ in Eq.~\eqref{eq:firstandsecondmomentsrhogaussianstate1} by determining the squeezing parameter and the average number of thermal particles from Eq.~\eqref{eq:xiandngaussianstatetatu1}. Then, the measured $p_n$ is compared to the photon distribution $q_n=\tau_{nn}$ obtained from the Fock expansion of $\hat\tau$ in Eq.~\eqref{eq:gaussiantargetstate1}. We define the deviation from Gaussianity as the squared Euclidian distance between $p_n$ and $q_n$, i.e.
\begin{equation}
Q\left(\hat\rho\right)\coloneqq\sum_{n=0}^{\infty}{\left(p_n-q_n\right)}^2 \,\, .
\label{eq:nongaussianityfromeuclidianentropy1}
\end{equation}
As discussed in Ref.~\cite{DAuria:05}, $p_n$ is expected to be close to $q_n$ below the oscillation threshold, while deviations from $q_n$ are observed as the threshold is approached, which motivates the choice of Eq.~\eqref{eq:nongaussianityfromeuclidianentropy1} as a measure of non-Gaussianity of the quantum state.

\begin{figure}[t]
\centering
\includegraphics[width=8.4cm]{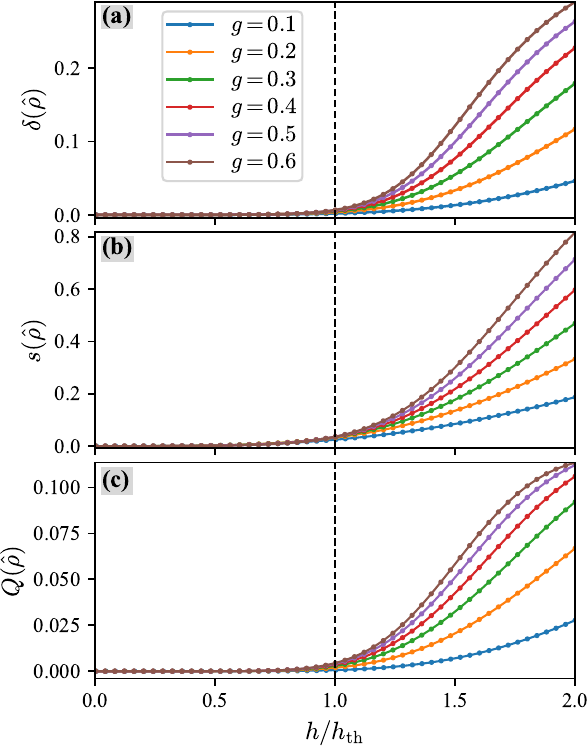}
\caption{Quantification of the non-Gaussian nature of the state $\hat\rho$. \textbf{(a)} Degree of non-Gaussianity $\delta(\hat\rho)$ from the Hilbert-Schmidt distance in Eq.~\eqref{eq:tracerhorhogaussian3}, \textbf{(b)} Quantum relative entropy $s(\hat\rho)$ from Eq.~\eqref{eq:measurenongaussianityentropy1}, and \textbf{(c)} Non-Gaussianity $Q(\hat\rho)$ from photon distribution in Eq.~\eqref{eq:nongaussianityfromeuclidianentropy1}. Data are shown as a function of the pump amplitude $h$ relative to the value of the classical oscillation threshold $h_{\rm th}=2g$, for different intrinsic loss parameters $g$ as in the legend. The vertical black dashed line marks the threshold value $h=h_{\rm th}$. Other numerical parameters are $n_{\rm max}=40$ and $\beta=0.1$.}
\label{fig:comparisondegreeofnongaussianity1}
\end{figure}

\subsection{Numerical results}
\label{sec:numericalresults1}
Figure~\ref{fig:comparisondegreeofnongaussianity1} shows the degree of non-Gaussianity from our numerical simulations, comparing $\delta(\hat\rho)$ from Eq.~\eqref{eq:tracerhorhogaussian3} in panel \textbf{(a)}, $s(\hat\rho)$ from Eq.~\eqref{eq:measurenongaussianityentropy1} in panel \textbf{(b)}, and $Q(\hat\rho)$ from Eq.~\eqref{eq:nongaussianityfromeuclidianentropy1} in panel \textbf{(c)}. Data are shown as a function of the pump amplitude $h$ relative to the classical threshold $h_{\rm th}$, which is marked in the plot as the vertical dashed black line. Different colors refer to different values of $g$ as in the legend. Other numerical parameters are $\beta=0.1$ and $n_{\rm max}=40$. Clearly, when truncating the Hilbert space, all quantities where the summation over the Fock states appears are evaluated by summing up to the Fock state with $n_{\rm max}-1$ particles.

As evident from the figure, all measured quantities show the same qualitative picture: They increase monotonously, being very close to zero below threshold and rapidly deviating from zero above threshold. In other words, the quantum state $\hat\rho$ is well approximated by a Gaussian state below threshold, while it becomes highly non-Gaussian above threshold. The fact that the curves at lower $g$ are below those at higher $g$ is a consequence of the fact that data are taken as a function of $h/h_{\rm th}$.

\begin{figure}[t]
\centering
\includegraphics[width=8.4cm]{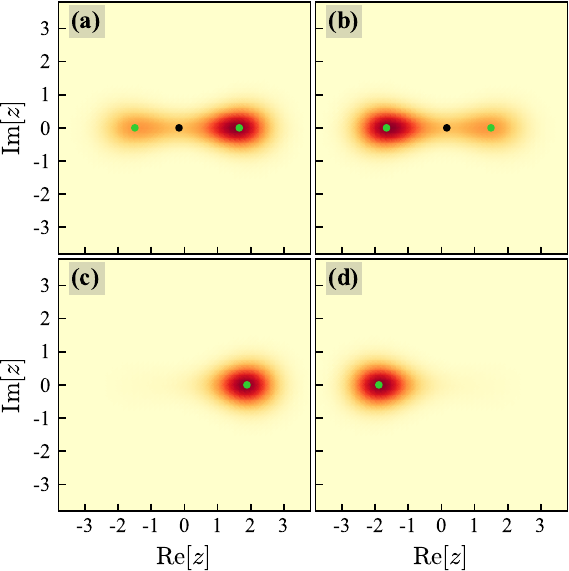}
\caption{Colormap of the Wigner function $W(z)$, as in Fig.~\ref{fig:wignerfunctioncomputation1}, in the presence of a real additive field $F$ as in Eq.~\eqref{eq:onephotonpumpfield1} with value \textbf{(a)} $F=0.02$, \textbf{(b)} $F=-0.02$, \textbf{(c)} $F=0.1$, and \textbf{(d)} $F=-0.1$. Numerical parameters are: $n_{\rm max}=40$, $g=0.5$, $\beta=0.1$, and $h=1.5$. Compared to Fig.~\ref{fig:wignerfunctioncomputation1}, which is for $F=0$, a nonzero $F$ breaks the inversion symmetry in phase space, resulting into an asymmetric $W(z)$, enhancing the lobe for ${\rm Re}[z]>0$ or ${\rm Re}[z]<0$ when $F>0$ or $F<0$, respectively, while suppressing the opposite one as $|F|$ increases.}
\label{fig:wignerfunctioncomputation2}
\end{figure}

We remark that, in our numerical simulations, the computation of $s(\hat\rho)$ in Eq.~\eqref{eq:measurenongaussianityentropy1} is significantly less demanding compared to that of $\delta(\rho)$ and $Q(\hat\rho)$ in Eqs.~\eqref{eq:tracerhorhogaussian3} and~\eqref{eq:nongaussianityfromeuclidianentropy1}, respectively. This is because $\delta(\hat\rho)$ and $Q(\hat\rho)$ require the computation of both $\hat\rho$ and $\hat\tau$ in the Fock basis. In fact, determining $\tau_{mn}$ as Eq.~\eqref{eq:gaussiantargetstate1} requires to perform at least $n_{\rm max}^3$ numerical operations (which become $n_{\rm max}^2$ when only the diagonal elements of $\hat\tau$ are needed) when $\alpha=0$ in Eq.~\eqref{eq:firstandsecondmomentsrhogaussianstate1}. In the general case, when $\alpha\neq0$, the number of operations to determine $\tau_{mn}$ increase to $n_{\rm max}^5$ (the additional $n^2_{\rm max}$ operations come from the displacement operator).

In addition to this, the calculation of $\tau_{mn}$ is strongly affected by the truncation of the Hilbert space (because the unitarity of the displacement and squeezing operators, as well as the proper normalization of the thermal state, are strictly speaking found only when the Hilbert space has infinite dimension), and therefore the numerical calculation of $\delta(\hat\rho)$ and $Q(\hat\rho)$ intrinsically carries with it an additional source of truncation error. This additional truncation error is reduced by increasing $n_{\rm max}$ until no sensitive change in the numerical results is observed. In our numerics, we indeed checked that no sensitive change of data occurred for $n_{\rm max}>40$. Instead, computing $s(\hat\rho)$ needs only $\hat\rho$, since also $\overline{n}$ in Eq.~\eqref{eq:measurenongaussianityentropy3} is found from the covariance matrix of $\hat\rho$ as in Eq.~\eqref{eq:xiandngaussianstatetatu1}, which makes its computation not only less demanding but also more accurate compared to the other two metrics shown in Fig.~\ref{fig:comparisondegreeofnongaussianity1}.

\section{Inclusion of an additive field}
\label{sec:inclustionofadditivefield1}
In this section, we analyze the non-Gaussianity of the quantum state by including an additive field. This is done by adding to the parametric gain Hamiltonian in Eq.~\eqref{eq:masterequationquantumopos2} the one-photon field
\begin{equation}
\hat H_F=iF\left(\hat a^\dag-\hat a\right) \,\, ,
\label{eq:onephotonpumpfield1}
\end{equation}
where $F\in\mathbb{R}$ quantifies the external field strength. The additional terms in the Lindbladian tensor in Eq.~\eqref{eq:masterequationquantumopos7} due to the presence of $\hat H_F$ are reported in Eq~\eqref{eq:nonzeroliuvillianfockbasis4} of Appendix~\ref{appendix:liuvilliantensorinthefockbasis1}.

\begin{figure}[t]
\centering
\includegraphics[width=8.4cm]{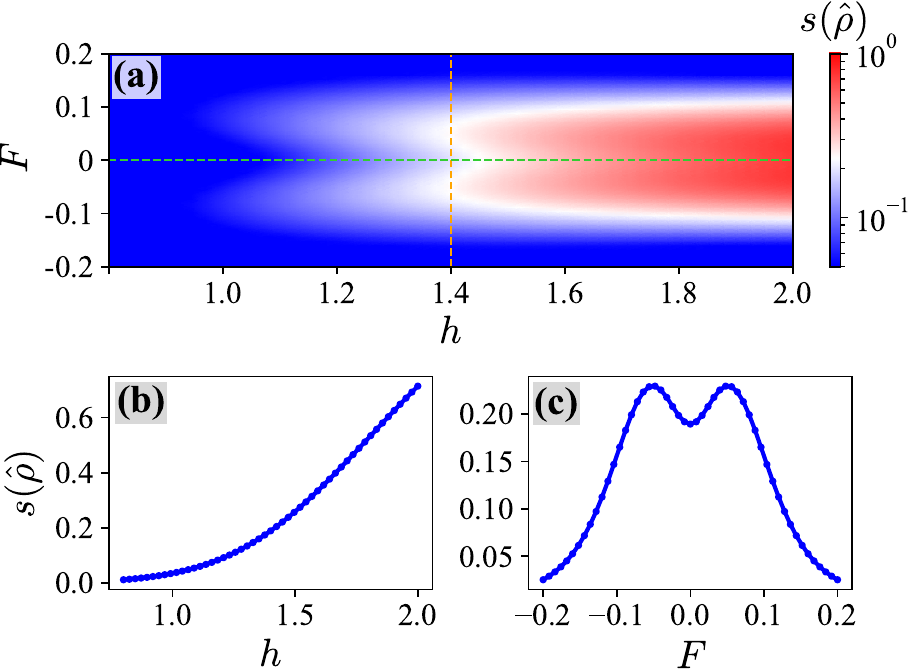}
\caption{Analysis of the quantum relative entropy $s(\hat\rho)$ as a function of the applied field ($F$) and pump amplitude ($h$). \textbf{(a)} Colormap of $s(\hat\rho)$ in the $F$ vs. $h$ plane. Color coding as in the color bar (note the logarithmic scale of the colormap). \textbf{(b)} Horizontal cut of the colormap at $F=0$, marked by the horizontal green dashed line in panel \textbf{(a)}, showing the monotonous increase as in Fig.~\ref{fig:comparisondegreeofnongaussianity1}. \textbf{(c)} Vertical cut of the colormap at $h=1.4$, marked by the horizontal orange dashed line in panel \textbf{(a)}, showing a symmetric structure peaked at $|F|>0$ and rapidly decreasing with $|F|$ (thus a non-monotonous behaviour). Other numerical parameters are: $n_{\rm max}=40$, $g=0.5$, and $\beta=0.1$.}
\label{fig:relativeentropycolomapandcuts1}
\end{figure}

In the adjoint master equation in Eq.~\eqref{eq:masterequationquantumopos4} and in its classical limit in Eq.~\eqref{eq:masterequationquantumopos5}, the applied field $\hat H_F$ in Eq.~\eqref{eq:onephotonpumpfield1} adds the extra term $F$ to the right-hand sides, i.e., a term that is not multiplied by $\hat a$ or $A$, respectively. This kind of additive field is relevant to Ising machines because it is envisioned to simulate applied fields fully optically in the simulated Ising model (see Ref.~\cite{eichler2023} for a recent work in an electronic oscillator network) without the need of electronic feedback mechanisms~\cite{PhysRevApplied.13.054059}, therefore preserving the quantum nature of the state.

The presence of the applied field breaks the inversion symmetry (i.e., polarizes the system) in phase space $\mathbf{R}\rightarrow-\mathbf{R}$, which is manifest by looking at the Wigner function and the classical fixed points of the equations of motion in Fig.~\ref{fig:wignerfunctioncomputation2}. As evident from the figure, the Wigner function loses its symmetric two-lobe structure found in Fig.~\ref{fig:wignerfunctioncomputation1}, which is for $F=0$. In particular, a positive (negative) $F$ enhances the lobe at ${\rm Re}[z]>0$ (${\rm Re}[z]<0$) in the complex plane, while suppressing the opposite one. In terms of the classical fixed points, the saddle (black, which corresponds to the origin for $F=0$) gradually approaches the attractor (green) at ${\rm Re}[z]<0$ for $F>0$ (or ${\rm Re}[z]>0$ for $F<0$) as $|F|$ increases, until the two fixed points eventually collide and annihilate each other via a saddle-node bifurcation. After this bifurcation, only the attractor at ${\rm Re}[z]>0$ for $F>0$ (or ${\rm Re}[z]<0$ for $F<0$) is found. In this parameter regime, the system is fully polarized, deterministically converging to the only remaining attractor.

We therefore see that the pump amplitude $h$ and the applied field $F$ play two antagonistic roles. The former tends to stabilize a state with two (possibly symmetric) configurations, while the latter induces imbalance, eventually polarizing (phase locking) the system to a single configuration. A natural question is how the combined effect of $h$ and $F$ influences the non-Gaussian character of the quantum state.

To answer this question, we extend the analysis on the measurements of non-Gaussianity of Sec.~\ref{sec:measirementsnongaussianity1} to the case of $F\neq0$. Since a nonzero $F$ induces a displacement in phase space, the first moments $\langle\hat X\rangle$ and $\langle\hat P\rangle$ are now nonzero, which in turn implies that $\alpha\neq0$ in the target Gaussian state $\hat\tau$ in Eq.~\eqref{eq:firstandsecondmomentsrhogaussianstate1}. Following the discussion in Sec.~\ref{sec:numericalresults1}, to keep a reasonable numerical complexity of the problem, we here quantify non-Gaussianity solely from the quantum relative entropy $s(\hat\rho)$ in Eq.~\eqref{eq:measurenongaussianityentropy1}. This choice is further supported by Fig.~\ref{fig:comparisondegreeofnongaussianity1}, which shows that the relative entropy provides qualitatively the same information as the other two metrics.

The numerical result of $s(\hat\rho)$ for different values of $h$ and $F$ is shown as a colormap in panel \textbf{(a)} of Fig.~\ref{fig:relativeentropycolomapandcuts1}. Our numerical results highlight a nontrivial interplay between $h$ and $F$. Indeed, while we observe that increasing $h$ causes a monotonous growth of $s(\hat\rho)$ at any $F$, generalizing the result in Fig.~\ref{fig:comparisondegreeofnongaussianity1} for $F=0$, increasing $|F|$ causes instead the quantum relative entropy to vary in a non-monotonous way: Starting from $F=0$ (green dashed line in the figure), it first increases, reaching a maximum value for nonzero $|F|$, and then it rapidly decreases. This behaviour is exemplified in panel \textbf{(c)}, where a vertical cut of $s(\hat\rho)$ at fixed $h$ is shown. From this analysis, we conclude that the parametric gain tends to drive the system into a regime of emerging non-Gaussianity. On the contrary, increasing $F$ above a certain value restores the Gaussian nature of the state.

\section{Conclusions}
\label{sec:perspectives1}
In this paper, we provided an \emph{ab initio} detailed numerical analysis of the emergence of the non-Gaussianity in the steady state of the single quantum optical parametric oscillator (OPO). We modeled the dynamical evolution of the system by a Lindblad master equation, where the Hermitian part described two-photon gain (parametric amplification), and the dissipation accounted for one- and two-photon losses, quantifying the intrinsic loss and amplitude-saturation nonlinearity, respectively. The full steady-state density matrix of the system was found by exact diagonalization of the Liouvillian tensor, resulting from the projection of the master equation onto the Fock (number) basis.

We first showed the Wigner function for different values of pump amplitude, and then discussed the measurement of non-Gaussianity from the density matrix, comparing three different quantities: Degree of non-Gaussianity from the Hilbert-Schmidt distance, quantum relative entropy, and non-Gaussianity from the covariance matrix and photon number distribution. By scanning the pump amplitude from zero to twice the classical oscillation threshold value, we revealed that all measured quantities monotonically increase with the pump amplitude, being close to zero below threshold and rapidly increasing above threshold. This result provides a quantitative clear evidence of how the steady state of the quantum OPO deviates from Gaussianity close to threshold, and becomes highly non-Gaussian for large gain.

We then extended the calculation of the Wigner function and quantum relative entropy to the quantum OPO in the presence of an additive field (one-photon drive). Our numerics pointed out the nontrivial interplay between parametric pump and additive field. Specifically, rising the pump amplitude generates a monotonous growth of non-Gaussianity, while a nonzero field first causes non-Gaussianity to grow, and then gives raise to a steep decrease for increasing field strength, suggesting the restoration of the Gaussian nature of the state.

Our work opens the future perspective to study without approximation how the quantum properties of small OPO networks such as non-Gaussianity and quantum entanglement evolve for different parameter regimes. Indeed, even if the \emph{ab initio} method here used becomes exponentially more demanding for increasing number of OPOs, it is still usable for systems of few OPOs only. Previous studied reported on the presence of quantum correlations in OPO networks using phase-space methods like the positive $P$-representation~\cite{PhysRevA.55.3014,PhysRevA.102.062419,yamamoto2021nopo}. An interesting perspective is to compare previous results with those obtainable from our \emph{ab initio} method, as well as from lattice approaches similar to matrix-product-state or density-matrix-renormalization-group methods~\cite{SCHOLLWOCK201196}.

\begin{acknowledgements}
We thank Cristiano Ciuti, Simone Felicetti, and Jacopo Tosca for useful discussions. C.C. acknowledges support from CN1 Quantum PNRR MUR CN 0000013 HPC.
\end{acknowledgements}

\appendix

\section{Liouvillian tensor in the Fock basis}
\label{appendix:liuvilliantensorinthefockbasis1}
In this appendix, we explicitly report the expression of the nonzero elements of the Liuvillian superoperator $\mathcal{L}$ in Eq.~\eqref{eq:masterequationquantumopos1} projected in the Fock basis. By recalling that the action of the annihiliation and creation operators on the Fock states is $\hat a|n\rangle=\sqrt{n}|n-1\rangle$ and $\hat a^\dag|n\rangle=\sqrt{n+1}|n+1\rangle$, and the definition $\rho_{mn}=\langle m|\hat\rho|n\rangle$, one has the projected Hermitian term
\begin{eqnarray}
&&\frac{1}{i}\,\langle m|\left[\hat H_0,\hat\rho\right]|n\rangle\nonumber\\
&=&\frac{h}{8}\left(\sqrt{m(m-1)}\,\rho_{m-2,n}-\sqrt{(m+1)(m+2)}\,\rho_{m+2,n}\right.\nonumber\\
&&\left.\hspace{0.3cm}+\sqrt{n(n-1)}\,\rho_{m,n-2}-\sqrt{(n+1)(n+2)}\,\rho_{m,n+2}\right) \,\, . \nonumber\\
\label{eq:nonzeroliuvillianfockbasis1}
\end{eqnarray}
The projected one-photon dissipator in Eq.~\eqref{eq:masterequationquantumopos3} reads
\begin{eqnarray}
&&\langle m|\mathcal{D}_{\rm 1ph}\left(\hat\rho\right)|n\rangle\nonumber\\
&=&g\left(\sqrt{(m+1)(n+1)}\,\rho_{m+1,n+1}-\frac{m+n}{2}\,\rho_{mn}\right) \,\,,\nonumber\\
\label{eq:nonzeroliuvillianfockbasis2}
\end{eqnarray}
and the projected two-photon dissipator is
\begin{eqnarray}
&&\langle m|\mathcal{D}_{\rm 2ph}\left(\hat\rho\right)|n\rangle\nonumber\\
&=&\frac{\beta}{2}\sqrt{(m+1)(m+2)(n+1)(n+2)}\rho_{m+2,n+2}\nonumber\\
&&-\beta\frac{m(m-1)+n(n-1)}{4}\rho_{mn} \,\, .
\label{eq:nonzeroliuvillianfockbasis3}
\end{eqnarray}
Without additive field [i.e., $F=0$ in $\hat H_F$ in Eq.~\eqref{eq:onephotonpumpfield1}], the nonzero elements of $\mathcal{L}^{rs}_{mn}$ are therefore at $(r,s)=(m,n)$, $(m\pm2,n)$, $(m,n\pm2)$, $(m+1,n+1)$, and $(m+2,n+2)$, whose expression is retrieved from Eqs.~\eqref{eq:nonzeroliuvillianfockbasis1}-\eqref{eq:nonzeroliuvillianfockbasis3}. The inclusion of $F\neq0$ adds at the right-hand side of Eq.~\eqref{eq:masterequationquantumopos1} and therefore Eq.~\eqref{eq:masterequationquantumopos7} the term
\begin{eqnarray}
&&\frac{1}{i}\,\langle m|\left[\hat H_F,\hat\rho\right]|n\rangle\nonumber\\
&=&F\left(\sqrt{m}\,\rho_{m-1,n}-\sqrt{m+1}\,\rho_{m+1,n}\right.\nonumber\\
&&+\left.\sqrt{n}\,\rho_{m,n-1}-\sqrt{n+1}\,\rho_{m,n+1}\right) \,\, ,
\label{eq:nonzeroliuvillianfockbasis4}
\end{eqnarray}
therefore yielding other nonzero elements of $\mathcal{L}^{rs}_{mn}$ at $(r,s)=(m\pm1,n)$ and $(m,n\pm1)$. Before diagonalization, $\mathcal{L}^{rs}_{mn}$ is reshaped as a matrix $\mathcal{L}_{pq}$ where $p=m+n_{\rm max}n$ and $q=r+n_{\rm max}s$. It is seen from Eqs.~\eqref{eq:nonzeroliuvillianfockbasis1}-\eqref{eq:nonzeroliuvillianfockbasis3} that $\mathcal{L}_{pq}$ is a very sparse matrix, with density of nonzero elements scaling as $1/n_{\rm max}^2$.

\section{Displacement operator in the Fock basis}
\label{appendix:displaementoperatorfockbasis1}
The matrix representation of the displacement operator $\hat D_z=e^{z\,\hat a^\dag-z^*\,\hat a}$ in the Fock basis follows from the fact that $\hat a|n\rangle=\sqrt{n}|n-1\rangle$ and $\hat a^\dag|n\rangle=\sqrt{n+1}|n+1\rangle$, and from Baker-Campbell-Housdorff theorem, which allows to write $\hat D_z=e^{z\,\hat a^\dag-z^*\,\hat a}=e^{-{|z|}^2/2}\,e^{z\,\hat a^\dag}\,e^{-z^*\,\hat a}$. For $m\geq n$, one can explicitly compute the matrix element
\begin{equation}
\langle n|\hat D_z|m\rangle=\sqrt{\cfrac{n!}{m!}}\,e^{-{|z|}^2/2}{(-z^*)}^{m-n}L^{(m-n)}_n\left({|z|}^2\right) \,\, ,
\label{eq:displacementoperatorinthefockbasis1}
\end{equation}
where $L^{(\alpha)}_n(x)$ is the generalized Laguerre polynomial~\cite{zwillinger2002crc}. The element for $m<n$ is found by using the fact that $\hat D_z^\dag=\hat D_{-z}$, i.e., $\langle n|\hat D_z|m\rangle={(\langle m|\hat D^\dag_z|n\rangle)}^*={(\langle m|\hat D_{-z}|n\rangle)}^*$, and therefore one has for $m<n$
\begin{equation}
\langle n|\hat D_z|m\rangle=\sqrt{\frac{m!}{n!}}\,e^{-{|z|}^2/2}z^{n-m}L^{(n-m)}_m\left({|z|}^2\right) \,\, .
\label{eq:displacementoperatorinthefockbasis2}
\end{equation}

\section{Covariance matrix and purity of the squeezed thermal state}
\label{appendix:covariancematrixsqueezedthermalstate1}
In this appendix, we recall the expression of the covariance matrix $\mathbf{\Sigma}_{\rm G}$ and purity of the displaced squeezed thermal state $\hat\tau=\hat D_\alpha\,\hat S(\xi)\,\hat\rho_{\rm th}(\overline{n})\,\hat S^\dag(\xi)\,\hat D^\dag_\alpha$ in Eq.~\eqref{eq:firstandsecondmomentsrhogaussianstate1}, where $\hat D_\alpha=e^{\alpha\,\hat a^\dag-\alpha^*\,\hat a}$ and $\hat S(\xi)=e^{\left(\xi^*\,\hat a\hat a-\xi\,\hat a^\dag\hat a^\dag\right)/2}$, and $\hat\rho_{\rm th}(\overline{n})$ is as in Eq.~\eqref{eq:fockbasisexpansionthermalstate1}. As recalled in Sec.~\ref{sec:determinationofthegaussianreferencestate1}, the covariance matrix of $\hat\tau$ is unaffected by the displacement $\hat D_\alpha$. Let us define for simplicity $\xi=r\,e^{i\varphi}$ in terms of its absolute value $r=|\xi|$ and phase $\varphi={\rm arg}(\xi)$. First, one recalls that the covariance matrix $\mathbf{\Sigma}_{\rm sqv}(r,\varphi)$ of the squeezed vacuum state $\hat S(\xi)|0\rangle\langle0|\hat S^\dag(\xi)$ is given by $\mathbf{\Sigma}_{\rm sqv}(r,\varphi)=\mathcal{R}(\varphi/2)\,\mathbf{\Sigma}_{\rm sqv}(r,0)\,\mathcal{R}^T(\varphi/2)$ where  $\mathcal{R}(\phi)=\bigl( \begin{smallmatrix}\cos(\phi) & -\sin(\phi)\\ \sin(\phi) & \cos(\phi)\end{smallmatrix}\bigr)$ is the rotation matrix, $\mathbf{\Sigma}_{\rm sqv}(r,0)=\frac{1}{2}\,{\rm diag}(e^{-2r},e^{2r})$, and $T$ denotes the transposition. The covariance matrix of the squeezed thermal state readily follows: $\mathbf{\Sigma}_{\rm G}=(2\overline{n}+1)\mathbf{\Sigma}_{\rm sqv}(r,\varphi)$.

Since the displacement and squeezing operator are unitary and the trace is cyclic, the purity of $\hat\tau$ reduces to the purity of the thermal state, i.e., ${\rm Tr}[\hat\tau^2]={\rm Tr}[\hat\rho^2_{\rm th}(\overline{n})]=\sum_{n=0}^{\infty}f_n^2=1/(2\overline{n}+1)$.

\section{Squeezing operator in the Fock basis}
\label{appendix:squeezingoperatorfockbasis1}
In this appendix, we report for the sake of completeness the explicit expression of the matrix representation of the squeezing operator $\hat S(\xi)=e^{\left(\xi^*\,\hat a\hat a-\xi\,\hat a^\dag\hat a^\dag\right)/2}$ with $\xi=r\,e^{i\varphi}$ in the Fock basis. This is $\langle n|\hat S(\xi)|m\rangle=0$ for $m$ and $n$ with opposite parity, while for $m$ and $n$ of the same parity one has
\begin{widetext}
\begin{equation}
\langle n|\hat S(\xi)|m\rangle=\left\{
\begin{array}{ll}
\displaystyle{\left(-\frac{\zeta}{2}\right)}^{(n-m)/2}e^{-(\eta/2)(m+1/2)}\sqrt{n!\,m!}\,\sum_{k=0}^{\lfloor m/2\rfloor}{\left(-\frac{{|\zeta|}^2e^\eta}{4}\right)}^k\frac{1}{(m-2k)!\,k!\,[(n-m)/2+k]!}  & \quad (n\geq m)\\\\
\displaystyle{{\left(\frac{\zeta^*}{2}\right)}^{(m-n)/2}e^{-(\eta/2)(n+1/2)}\sqrt{m!\,n!}\,\sum_{k=0}^{\lfloor n/2\rfloor}{\left(-\frac{{|\zeta|}^2e^\eta}{4}\right)}^k\frac{1}{(n-2k)!\,k!\,[(m-n)/2+k]!}} & \quad (n< m)
\end{array}
\right. \,\, ,
\label{eq:squeezingoperatorfockstatesprop1}
\end{equation}
\end{widetext}
where $\zeta=e^{i\varphi}\,\tanh(r)$ and $\eta=2\,\log[\cosh(r)]$, and $\lfloor\cdot\rfloor$ is the floor function. This result is derived after a chain of identities first by using the operator ordering of $\hat S(\xi)$~\cite{barnett2002methods}, and then by using $\hat a|n\rangle=\sqrt{n}|n-1\rangle$ and $\hat a^\dag|n\rangle=\sqrt{n+1}|n+1\rangle$, similar to Appendix~\ref{appendix:displaementoperatorfockbasis1}. The explicit calculation can be also found in Ref.~\cite{Varro_2022}.



%

\end{document}